\documentclass{elsart}

\usepackage{epsfig}

\usepackage{amssymb}

\journal{New Astronomy Reviews}

\begin{document}

\begin{frontmatter}



\title{The MEGA Advanced Compton Telescope Project}


\author[MPE]{P. F. Bloser},
\ead{bloser@mpe.mpg.de}
\author[MPE]{R. Andritschke},
\author[MPE]{G. Kanbach},
\author[MPE]{V. Sch\"onfelder},
\author[MPE]{F. Schopper},
\author[MPE]{A. Zoglauer},
\author[MPE,coll]{for the MEGA Collaboration}

\address[MPE]{Max-Planck-Institut f\"ur extraterrestrische 
Physik, Giessenbachstrasse, D-85748 Garching, Germany}

\address[coll]{France: CESR, Toulouse; USA: UNH, Durham, NH; NASA/GSFC,
Greenbelt, MD; NRL, Washington, D.C.; Columbia University, NY, NY; 
IGPP, UCR, Riverside, CA; UA, Huntsville, AL}

\begin{abstract}
The goal of the Medium Energy Gamma-ray Astronomy (MEGA) telescope is 
to improve sensitivity at medium gamma-ray energies
(0.4--50 MeV) by at least an order of magnitude over that of COMPTEL.
This will be achieved with a new compact design that allows for a very 
wide field of view, permitting a sensitive all-sky survey and the 
monitoring of transient and variable sources.  The key science objectives
for MEGA include the investigation of cosmic high-energy particle 
accelerators, studies of nucleosynthesis sites using gamma-ray lines, 
and determination of the large-scale structure of galactic and cosmic 
diffuse background emission.  MEGA records and images gamma-ray events
by completely tracking both Compton and pair creation interactions in 
a tracker of double-sided silicon strip detectors and a calorimeter of
CsI crystals able to resolve in three dimensions.  We present initial
laboratory calibration results from a small prototype MEGA telescope.

\end{abstract}

\end{frontmatter}


\section{Introduction}
\label{sec-intro}

Since the demise of the COMPTEL instrument on CGRO, the 
medium energy gamma-ray region of the spectrum (0.4--50 MeV) has been
left unattended and will not be adequately covered by currently-planned
missions.  While the INTEGRAL mission (2002) will provide pointed observations 
of line and continuum sources up to 10 MeV and the AGILE ($\sim 2003$) and 
GLAST ($\sim 2006$) missions promise great advances at energies 
above 100 MeV, an Advanced Compton Telescope (ACT) mission, with all-sky
survey capabilities and a narrow line sensitivity of $10^{-7}$ photons 
cm$^{-2}$ s$^{-1}$ (a factor of 100 better than COMPTEL), lies at least a 
decade in the future.  
\begin{figure}[t]
\begin{minipage}[t]{6.75cm}
\epsfig{figure=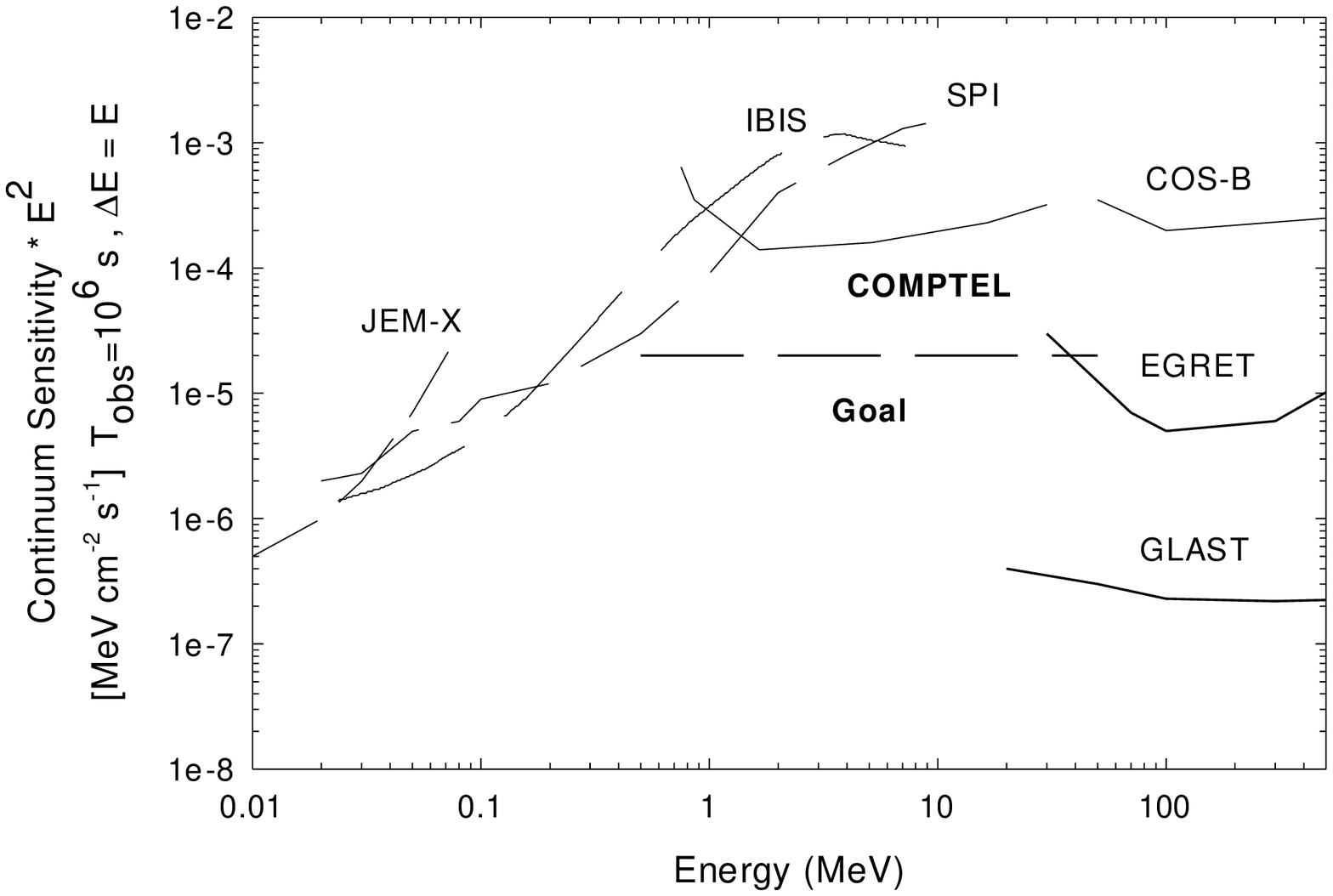,height=5.5cm,width=6.75cm}
\caption{Sensitivity levels for continuum point sources for various past
and future missions.  The goal for MEGA is shown by the dashed line.}
\label{fig-goal}
\end{minipage}
\hspace*{0.3cm}
\begin{minipage}[t]{6.75cm}
\begin{center}
\epsfig{figure=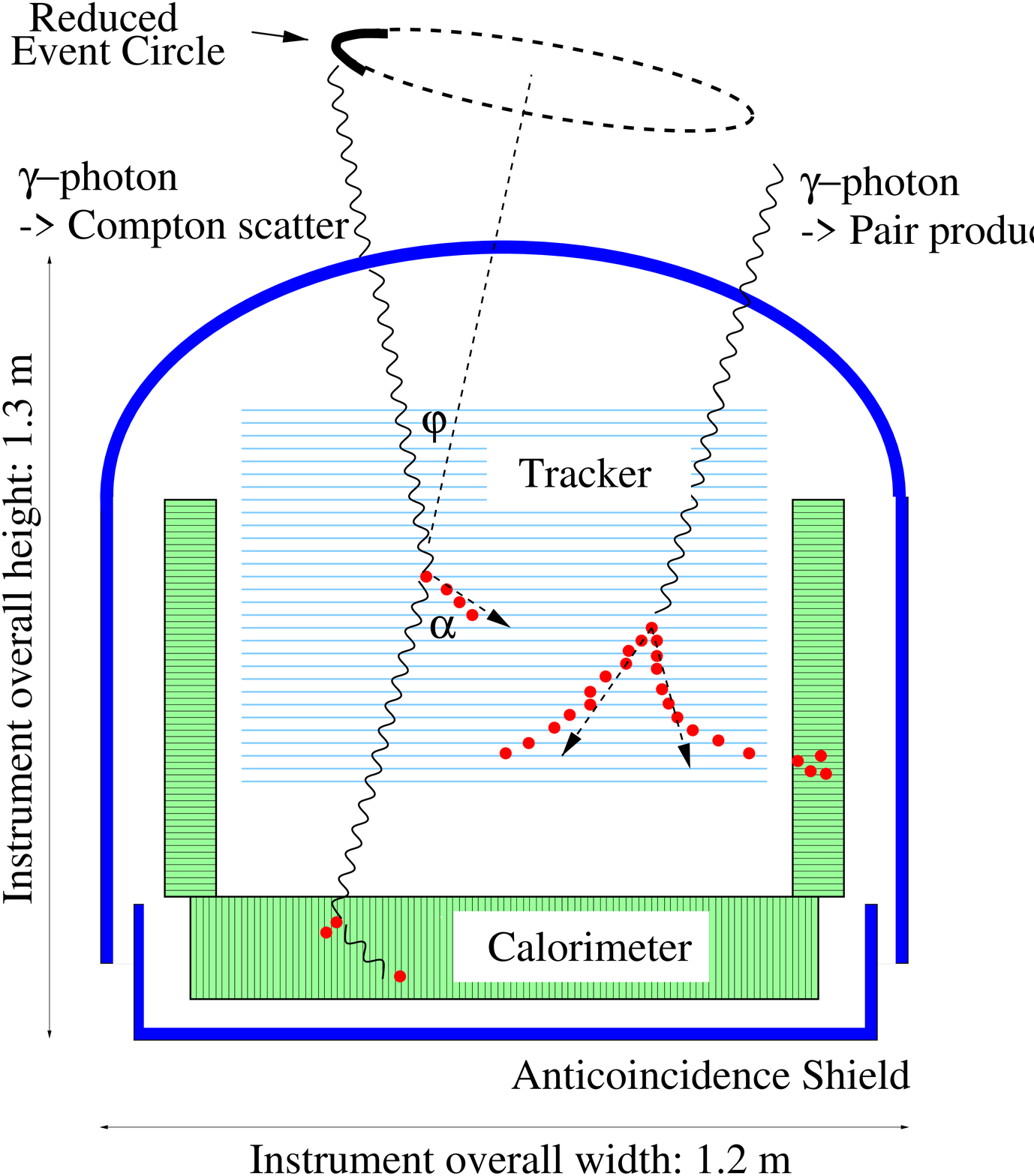,height=5.5cm}
\end{center}
\caption{Schematic of the MEGA design, showing tracker and calorimeter.  Both
Compton scatters and pair production events can be recorded.}
\label{fig-principle}
\end{minipage}
\end{figure}
Figure~\ref{fig-goal} shows the sensitivities of past and planned gamma-ray 
missions.  It is obvious that, at high energies, technological advances have
produced three ``generations'' of instruments (COS-B, EGRET, and GLAST), each
with a factor of $\sim 10$ improvement in sensitivity.  In a similar fashion, 
using technology that is available today,
a second-generation medium energy mission could provide an intermediate
step toward ACT and improve the sensitivity of
COMPTEL by a factor of 10 (indicated by the dashed line in Figure~\ref{fig-goal}).
Beginning this project now would allow a launch in 
$\sim 2006$ and permit simultaneous observations with GLAST;
such broadband spectral coverage was shown by CGRO to be invaluable.  The
Medium Energy Gamma-ray Astronomy (MEGA) project at MPE aims to provide this
intermediate step.

\section{Scientific Objectives}
\label{sec-science}

The medium energy gamma-ray band is of particular importance for a wide
variety of important high-energy astrophysical problems, specifically those
involving nuclear, non-thermal, and relativistic processes.  The production of
elements in massive stars and supernovae results in unstable isotopes, such
as $^{26}$Al and $^{44}$Ti, that produce radioactive decay lines in the MeV
range.  Compact objects, including black hole binaries and gamma-ray pulsars,
possess Doppler-shifted nuclear de-excitation lines as well as continuum
spectral slopes, breaks, and cutoffs in the gamma-ray band that are characteristic 
of relativistic
particle acceleration and are highly variable.  Outside the galaxy, active 
galactic nuclei (AGN) are 
prodigious producers of gamma-rays, some especially so in the MeV range.  
These sources are also highly variable, and therefore must be monitored constantly.  
Continuous monitoring is also needed to study gamma-ray bursts, which produce
the bulk of their energy around 1 MeV.  In addition to such discrete sources,
both galactic and extra-galactic diffuse gamma-ray emission exists as well.  The
former arises from both cosmic-ray interactions in the interstellar medium and
radioactive material ejected from supernovae; the latter is likely the 
superposition of distant AGN.  The combined instruments of CGRO provided a 
first taste of all these phenomena, but many were only marginally detected.
For example, $^{56}$Co line emission from SN1991T, the only supernova Type Ia possibly
seen in gamma-rays,  was detected by COMPTEL at the
3--4 $\sigma$ level (Morris et al. 1998), and about half of the 10 ``MeV blazars'' 
are also seen near COMPTEL's detection threshold.
Larger samples with higher significance of all these phenomena, together with
continuous monitoring, are needed to learn more.

\section{Instrument Design}
\label{sec-instrument}

Achieving the desired improvment in sensitivity for an instrument with good
imaging and spectroscopy abilities in the MeV range is a difficult task.  At these
energies the overall interaction cross section goes through a minimum, and the
primary photon interaction produces long-range secondary photons and/or
electrons.  At about 8 MeV 
(in silicon, for example) the most likely interaction changes from Compton 
scattering to pair production.  In order to make full use of all interacting photons
one then requires both sufficient material depth to achieve high detection
efficiency and good spatial and energy resolution to track the secondary 
particles.  Solutions to this problem that are available with today's
technology include Xenon time projection
chambers (Aprile et al. 2000), stacks of germanium detectors (Boggs \& Jean 2000),
and combinations of solid-state detectors and scintillators (O'Neill et al. 2000).  

The basic MEGA telescope design in shown in Figure~\ref{fig-principle}.  It
consists of a {\em tracker}, in which the primary Compton scatter or pair creation
event takes place, and a {\em calorimeter}, which absorbs and measures the secondary
particles.  The tracker contains 32 layers of double-sided Si strip detectors
($6 \times 6$ wafers of 6 cm $\times$ 6 cm each, 500 $\mu$m thick, with a pitch
of 470 $\mu$m).  The tracker is enclosed on five sides by the calorimeter, made of
CsI cells (5 mm $\times$ 5 mm cross section, 8 cm long on the bottom, 4 cm long
on the sides) read out by Si PIN diodes.  For incident 
photon energies above 
$\sim 1.5$ MeV the Compton recoil electron in most events receives enough 
energy to be tracked
through multiple silicon layers; the origin of the photon can then be constrained
to lie on only a small segment of the standard Compton event circle (``reduced
event circle'' in Figure~\ref{fig-principle}), reducing
background in crowded fields.  Electron-positron pairs may be similarly tracked.  
The locations and energy deposits of all interactions may be 
analyzed to reconstruct a most likely sequence for each event, which leads to 
very efficient suppression of background events.  The entire MEGA assembly is
further surrounded by a plastic anticoincidence shield.

Simulations based on the MEGA prototype (see Section~\ref{sec-prototype}) 
using GEANT3 indicate
an effective area of about 100 cm$^2$ at 2 MeV for the full MEGA telescope, 
with an angular
resolution of $\sim 4^{\circ}$ FWHM and an energy resolution of $\sim 8$\% FWHM
at 2 MeV.  The field of view is quite large, $\sim 130^{\circ}$ FWHM, allowing all-sky
survey operations.  Replacing the PIN diodes in the calorimeter with low-noise
silicon drift diodes could improve the angular resolution to $\sim 2^{\circ}$
and the energy resolution to $\sim 4$\%.
An exciting prospect for an instument based on Compton scattering 
through large angles is high sensitivity to polarization.  Our calculations 
indicate that MEGA should be sensitive
to polarization levels of $\sim 15$\% for energies less than 1 MeV from a 
Crab-like source after 100 hours of observation.

\section{Mission Concept and Expected Results}
\label{sec-mission}

The baseline MEGA experiment has been considered in a pre-phase A study for a 
small satellite mission.  Based on this study, the detector will have a mass 
of $\sim 650$ kg and
dimensions of 1.3 m diameter by 1.1 m length.  Placed on a standard small 
satellite platform, the launch payload mass is about 950 kg with a diameter of
2 m and a length of 2.4 m. The electrical power requirement is $\sim 400$ W, and
the average telemetry rate 50 kbit s$^{-1}$.  The development time of MEGA
to launch would be about 5.5 years, after which a mission lifetime of 3--5 years 
is forseen.  This would allow considerable overlap with the GLAST mission, a 
highly desirable goal that should be pursued if at all possible.  MEGA would be
placed in a low earth orbit (500-550 km) in as low an inclination as possible in
order to provide a low-background environment.  MEGA would be operated in a
zenith-pointing mode, performing a continuous all-sky survey.  The wide field of
view provides a nearly complete scan with each orbit and allows the constant
monitoring of variable sources.  Real-time satellite telemetry is planned through
the TDRSS Demand Access System.  The data will be analyzed promptly for bursts
and transients, and appropriate alert messages will be sent to initiate follow-up
observations.

\begin{figure}[t]
\begin{center}
\begin{tabular}{lcr}
\hspace{-.5cm}\epsfig{figure=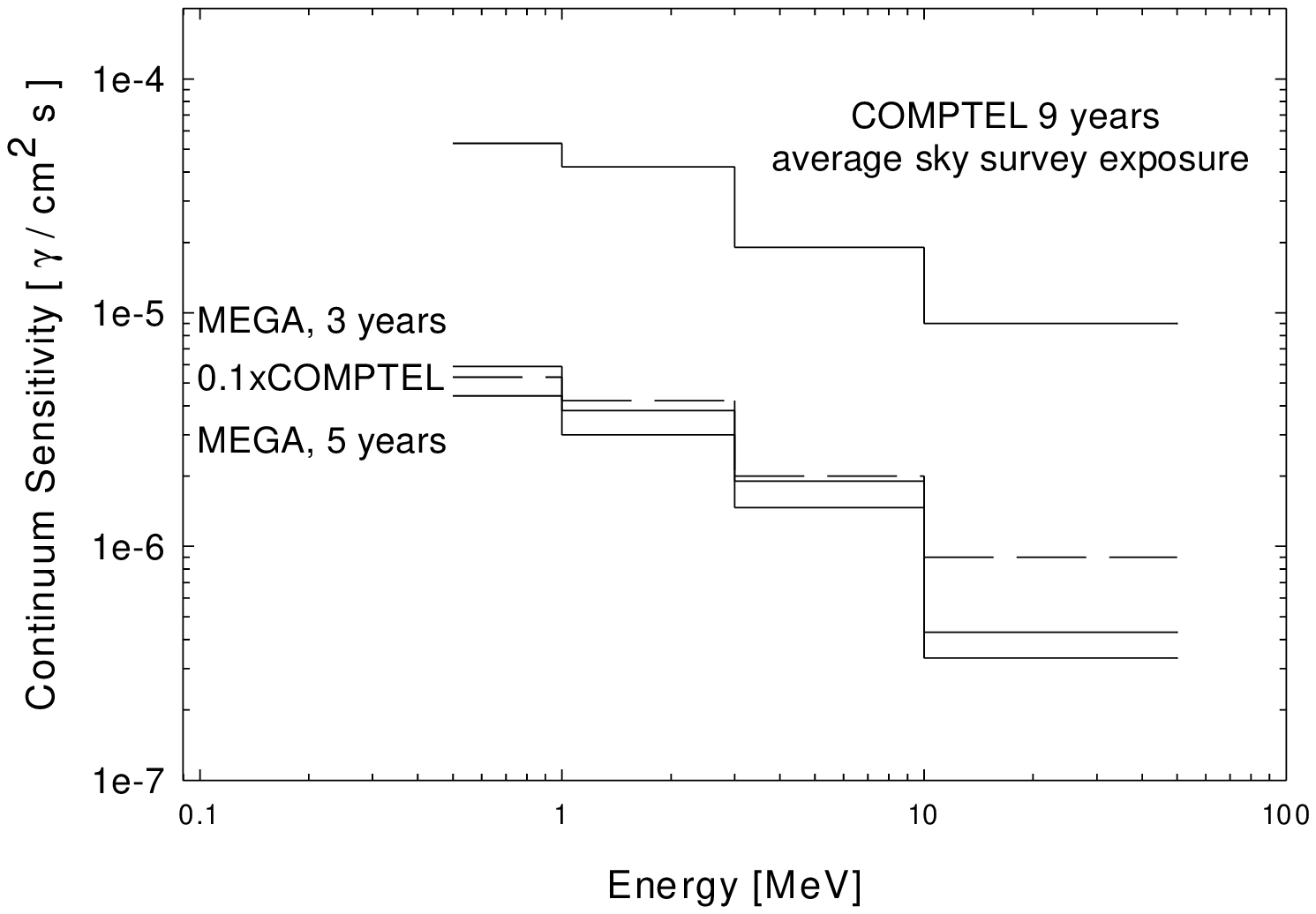,height=5.5cm,width=6.4cm} & &
\epsfig{figure=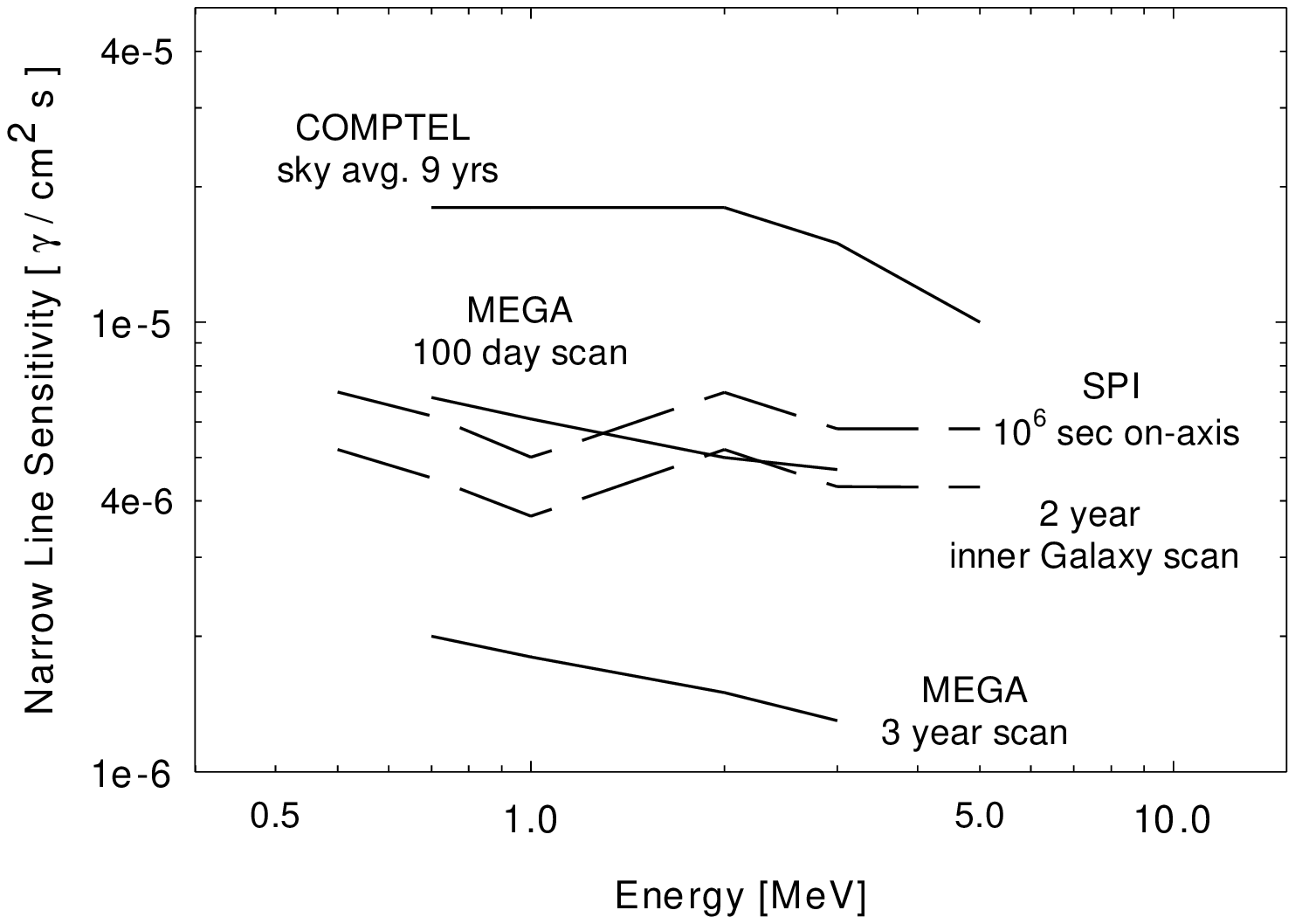,height=5.5cm,width=6.4cm}
\end{tabular}
\end{center}
\caption{Left: Sensitivity ($3\sigma$) for MEGA for sources with continuous 
spectra for 3 and
5 year surveys. The dashed line shows a 10-fold improvement over COMPTEL.  Right:
Sensitivity (also $3\sigma$) for narrow spectral lines, compared to COMPTEL and to 
SPI on INTEGRAL.}
\label{fig-sens}
\end{figure}

The survey sensitivity of MEGA has been derived from simulations by
calculating the relative number of source and background counts falling within
a circular area on the sky with width $3\sigma$ of the angular resolution.
Figure~\ref{fig-sens} shows the results for continuum and narrow line sources.  
The MEGA in-orbit background was assumed to be 
equivalent to 3 times the in-flight spectrum recorded by COMPTEL at 5 GV cutoff
rigidity, including events that appear as gamma-rays but are 
in fact caused by local background (concentrated in a bump
in the nuclear line region of 1--8 MeV).  
This background assumption will have to be checked carefully once
the instrument and satellite mass model and orbit have been precisely defined.  

A preliminary estimate of the number of sources detectable by MEGA in a 3-year
survey predicts that, within the galaxy, about 100 unidentified EGRET sources
should be seen.  The number of pulsars observed at MeV energies should grow to
about 10, and about a dozen black hole binaries like Cyg X-1 should be detected.
Although exact numbers are highly model-dependent,
we also expect to discover several young supernova remnants through their
$^{44}$Ti line emission, while a few novae should be detectable through
their lines each year.
Outside the galaxy, about 100 blazars and more than 10 radio and Seyfert galaxies
should be visible.  A gamma-ray burst will occur and be imaged in the large field
of view of MEGA every $\sim 2$ days.  We should also see 2--3 supernovae per year.
Far more detailed mapping of the spatial and spectral distribution of diffuse
emission, both lines and continuum, will be possible than was the case with 
COMPTEL.  Thus much larger samples will be available of all medium 
energy gamma-ray
astronomical objects, allowing the first glimpse provided by CGRO to be
expanded and explored.  The tantalizing marginal results of COMPTEL will
be unambiguously confirmed or denied, and science beyond mere detection will be
possible.

\section{Prototype Development}
\label{sec-prototype}

\begin{figure}[t]
\begin{minipage}[t]{6.75cm}
\epsfig{figure=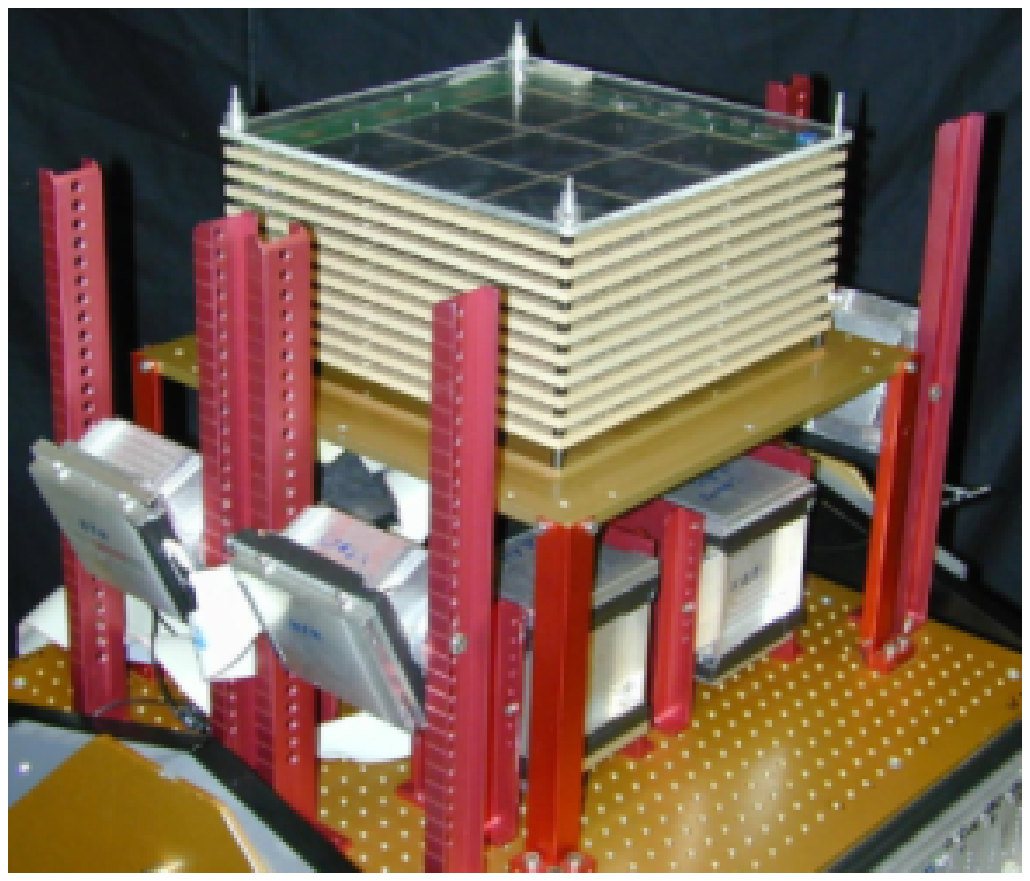,height=5.5cm,width=6.75cm} 
\caption{The MEGA prototype detector, with 10 tracker layers and 4 CsI blocks
visible.}
\label{fig-proto}
\end{minipage}
\hspace*{0.3cm}
\begin{minipage}[t]{6.75cm}
\begin{center}
\epsfig{figure=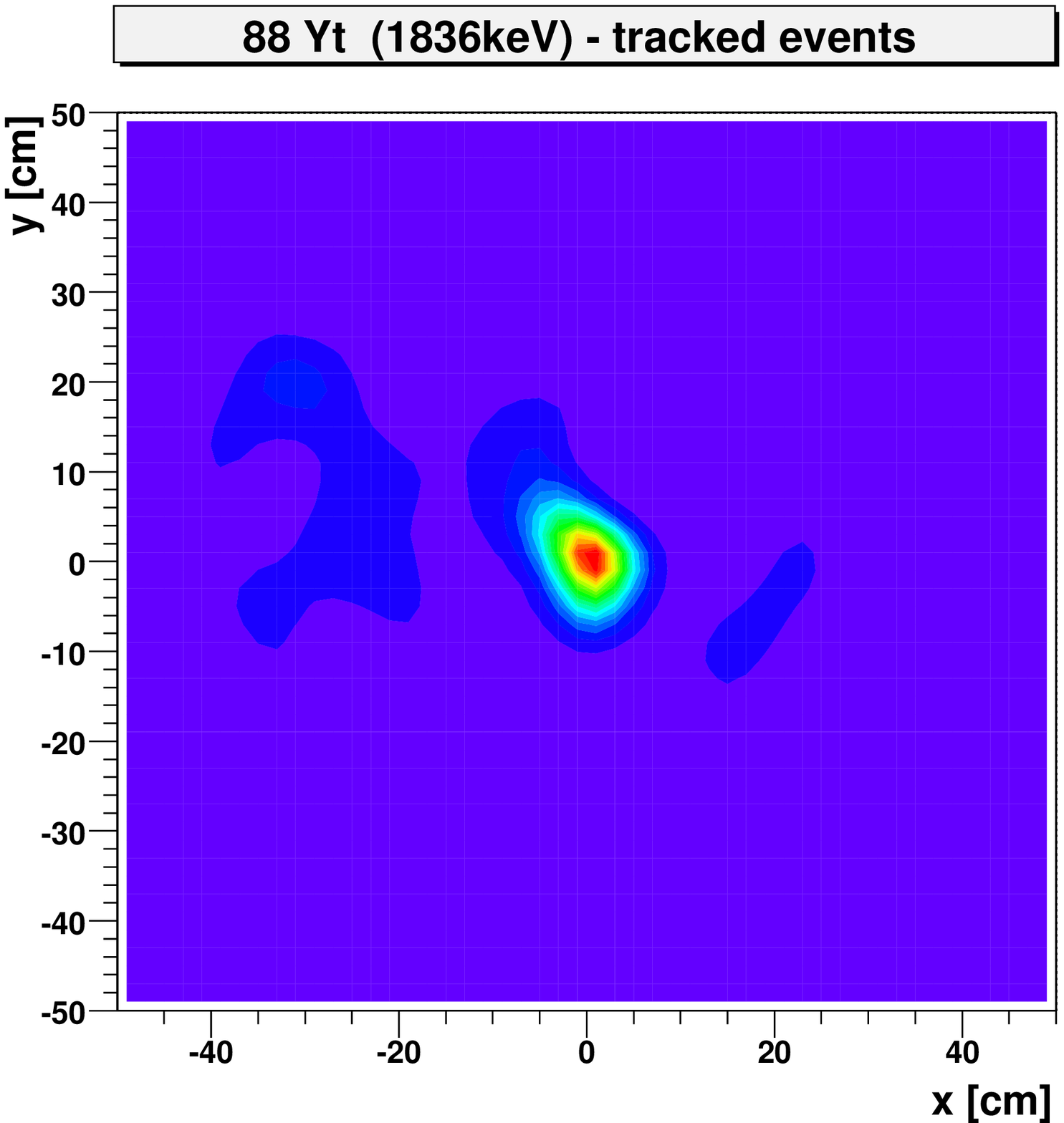,height=5.5cm,width=6.75cm}
\end{center}
\caption{Image of a $^{88}$Y source (1.836 MeV) located 80 cm from the prototype,
using tracked events only.}
\label{fig-image}
\end{minipage}
\end{figure}

A prototype of the MEGA telescope has been constructed in the 
laboratory at MPE, 
and is shown in Figure~\ref{fig-proto}. 
The prototype tracker has 10 layers of
$3 \times 3$ Si wafers each, and the calorimeter will comprise
20 modules of CsI detectors, each with a 10 $\times$ 12 array of crystals either
2 cm, 4 cm, or 8 cm deep (Schopper et al. 2000).  The 8 cm crystals have
PIN diodes on both ends, which allows the depth of the photon interaction
to be determined by the ratio of output light.  Both the tracker and calorimeter
detectors are read out by identical front-end ASIC chips (TA1.1 chip by IDE, 
self-triggering with
128 channels), custom-made front-end control units, and laboratory VME and NIM
electronics.  The individual detector units must be calibrated individually
using laboratory radioactive sources, and then the entire telescope may be
tested using higher energy sources and a coincident trigger.  Each recorded
event is reconstructed according to the most likely sequence of interactions
in the tracker (electrons) and calorimeter (scattered photons).  Images are
reconstructed using a list-mode maximum-likelihood expectation-maximization
method adapted from Wilderman et al. (1998).  Figure~\ref{fig-image} shows
an image of a $^{88}$Y source (1.836 MeV), located $\sim 80$ cm from the 
detector, constructed using only those events which produced tracks in the silicon
layers.  Only eight calorimeter modules were in place for this measurement.  
These results prove that the MEGA concept is viable and that the
detector technology and data analysis techniques are presently available to
construct a next-generation gamma-ray telescope.

In February 2002 the MEGA prototype will be taken to Duke University's High 
Intensity Gamma-ray Source for calibration measurements.  This facility 
produces a gamma-ray beam with tunable energy between 2 MeV and 55 MeV 
using Compton back-scattering
of UV laser photons in an electron storage ring.  The beam is highly polarized, 
providing an additional, valuable test of MEGA's sensitivity as a polarimeter.  
In June 2002
we plan to fly the MEGA prototype on a CNES balloon flight from Gap, France 
(in collaboration with CESR)
to measure background and test detector performance under flight conditions.  If
all goes well, the MEGA concept will have proven itself suitable for a 
space mission.

\end{document}